\documentclass[10pt]{article}

\usepackage{graphicx}

\begin{document}

\input amssym.tex

\title{On the de Sitter Tardyons and Tachyons}

\author{ION I. COTAESCU \footnote{e-mail: icotaescu@yahoo.com}\\{\it West University of Timi\c soara, V. Parvan Ave. 4,}\\
{\it Timi\c soara, RO-300223, Romania }}

\maketitle

\begin{abstract}
It is shown that on the de Sitter manifolds the tachyonic geodesics are restricted such that the classical tachyons cannot exist on this manifold at any time. On the contrary, the theory of the scalar quantum tachyons is free of any restriction. The tachyonic scalar and Dirac plane waves are deduced in this geometry, pointing out that these are well-defined, behaving as tempered distributions at any moment.

Pacs: 04.62.+v
\end{abstract}


\section{Introduction} 

In special relativity one knows three types of particles, tardyons (subluminal), light-like  and tachyons (superluminal).  The first two types of particles are of our world, inside the light-cone, while the tachyons seems to live in another one, outside the light-cone.  These two worlds seems to be completely separated as long as there are no direct physical evidences about the tardyon-tachyon interactions. For this reason the tachyons are the most attractive hypothetical	objects for speculating in some domain in physics where we have serious difficulties in building coherent theories. We mention, as an example, the presumed role of the tachyons in the early brane cosmology \cite{Sen}. However, here we do not intend to comment on this topics, restricting ourselves to analyze, from  the mathematical point of view, the possibility  to meet classical or quantum scalar or Dirac tachyons on the de Sitter backgrounds.

The de Sitter manifold, denoted from now by $M$, is local-Minkowskian such that the tachyons can be defined as in special relativity, with the difference that their properties are arising now from the specific high symmetry of the de Sitter manifolds. It is known that
 the isometry group of $M$, denoted by $I(M)=SO(1,4)$, is in fact the gauge group of the Minkowskian five-dimensional manifold $M^5$ embedding $M$. The unitary irreducible representations of the corresponding group  $S(M)={\rm Spin}(1,4)=Sp(2,2)$ are well-studied \cite{linrep} and used in various applications. Many authors exploited this high symmetry for building theories of quantum fields,  either by constructing symmetric two-point functions, avoiding thus the canonical quantization \cite{Wood}, or by using directly these unitary representations for finding field equations but without considering covariant representations  \cite{Gaz, Mos}. Another approach which applies the canonical quantization to the covariant fields transforming according to induced covariant representations was initiated by Nachtmann \cite{Nach}  many years ago and continued in few of our papers \cite{CNach,CCC,CQED}. 

In what follows we would like to study the tachyons on $M$ in this last framework by using the traditional definition of tachyons as particles of real squared masses but of opposite sigs with respect to the tadyonic ones. For this reason our results are different from other approaches \cite{Gaz, Mos} where  particles whose squared masses are  supposed to be  complex numbers are considered as tachyon \cite{Mos}.  Thus we find a result that seems to be somewhat paradoxical, namely that the classical tachyons cannot exist on $M$ at any time while the quantum scalar and Dirac particles behaves normally on this manifold, without any restriction.

The paper is organized as follows. In the next section we present the geodesics depending on a parameter giving their types (tardyonic, tachyonic or light-like) and we argue why the tachyonic lifetime is restricted on $M$. The next section is devoted to the quantum scalar and Dirac tachyons whose theory does not meet difficulties such that the mode functions are defined correctly as tempered distributions on the entire space at any moment. Finally, we present the principal conclusion concerning the discrepancy between the classical and quantum cases.

\section{Classical de Sitter geodesics}

The de Sitter spacetime $M$ is defined as the hyperboloid of radius $1/\omega$ \footnote{We denote by $\omega$
the Hubble de Sitter constant since  $H$ is reserved for the energy operator} in the five-dimensional flat spacetime $(M^5,\eta^5)$ of coordinates $z^A$  (labeled by the indices $A,\,B,...= 0,1,2,3,4$) and metric $\eta^5={\rm diag}(1,-1,-1,-1,-1)$. The local charts $\{x\}$  can be introduced on $(M,g)$ giving the set of functions $z^A(x)$ which solve the hyperboloid equation,
\begin{equation}\label{hip}
\eta^5_{AB}z^A(x) z^B(x)=-\frac{1}{\omega^2}\,.
\end{equation}
Here we use the chart $\{t,\vec{x}\}$ with the conformal time $t$ and Cartesian spaces coordinates $x^i$ defined by
\begin{eqnarray}
z^0(x)&=&-\frac{1}{2\omega^2 t}\left[1-\omega^2({t}^2 - \vec{x}^2)\right]
\nonumber\\
z^i(x)&=&-\frac{1}{\omega t}x^i \,, \label{Zx}\\
z^4(x)&=&-\frac{1}{2\omega^2 t}\left[1+\omega^2({t}^2 - \vec{x}^2)\right]
\nonumber
\end{eqnarray}
This chart  covers the expanding portion, $M_+$,  of $M$ for $t \in (-\infty,0)$
and $\vec{x}\in {\Bbb R}^3$ while the collapsing part, $M_-$,  is covered by
a similar chart but with $t >0$. Both these charts have the
conformal flat line element,
\begin{equation}\label{mconf}
ds^{2}=\eta^5_{AB}dz^A(x)dz^B(x)=\frac{1}{\omega^2 {t}^2}\left({dt}^{2}-d\vec{x}^2\right)\,.
\end{equation}
We remind the reader that on each portion one can introduce a FLRW chart with proper time defined as, 
\begin{equation}\label{ttpr}
t_{proper}=\left\{
\begin{array}{cllll}
-\frac{1}{\omega}\ln (-\omega t)&{\rm on} & M_+&{\rm  for}&-\infty<t<0\,,\\
&&&&\\
\frac{1}{\omega}\ln (\omega t)&{\rm on}& M_-&{\rm for} &0<t<\infty \,,
\end{array}\right. 
\end{equation}
such that  this spans the whole real axis,  $t_{proper}\in (-\infty,\infty)$, on each portion.

By definition,  the de Sitter spacetime $M$ is a homogeneous space of the
pseudo-orthogonal group $SO(1,4)$ which is in the same time the
gauge group of the metric $\eta^{5}$ and the isometry group, $I(M)$,
of $M$.  The classical conserved quantities as well as the  basis-generators
of the covariant representations of the isometry group can be calculated with the help of  the  Killing vectors  $k_{(AB)}$ which have   the following components:
\begin{eqnarray}
&&k^0_{(0i)}=k^0_{(4i)}=\omega t x^i\,,\qquad k^j_{(0i)}=k^j_{(4i)}-
\frac{1}{\omega}\,\delta^j_i= \omega x^i x^j-\delta^j_i\chi\,,\label{chichi1}\\
&&k^0_{(ij)}=0\,,\quad k^l_{(ij)}=\delta^l_jx^i-\delta^l_i
x^j\,;\qquad k^0_{(04)}=t\,,\quad
k^i_{(04)}=x^i\,.\label{chichi2}
\end{eqnarray}
where 
\begin{equation}
\chi=\frac{1}{2\omega}\left[1-\omega^2(t^2-{\vec{x}}^2)\right]\,.
\end{equation}

Furthermore, It is a simple exercise to integrate the geodesic equations and to
find the conserved quantities on a geodesic trajectory on $M$. Using the standard notation 
$u^{\mu}=\frac{dx^{\mu}(s)}{ds}$ we bear in mind that the principal invariant $u^2=g_{\mu\nu}u^{\mu}u^{\nu}=\epsilon$ along the geodesics $x=x(s)$  determines the type of this trajectory, i. e.  tardyonic ($\epsilon=1$), light-like ($\epsilon=0$) or tachyonic ($\epsilon=-1$). All the other conserved quantities along the geodesics are  proportional to $k_{(AB)\,\mu}u^{\mu}$ and can be derived by using the Killing vectors (\ref{chichi1}) and (\ref{chichi2}). 

We assume first  that in the chart $\{ t,\vec{x}\}$ the particle of mass $m\not=0$ has the
conserved momentum $\vec{p}$ of components $p\,^i=\omega m
(k_{(0i)\,\mu}-k_{(4i)\,\mu})u^{\mu}$ so that  we can write
\begin{equation}
u^0=\frac{dt}{ds}=-\omega t \sqrt{\epsilon+\frac{\omega^2{p}^{2}}{m^2}t^2}\,,
\qquad u^i=\frac{d{x^i}}{ds}=(\omega t)^2 \frac{p\,^i}{m}\,,
\end{equation}
using the notation $p=|{\vec{p}}\,|$. Hereby we deduce the
trajectory,
\begin{equation}
{x}^i(t)={x}_0^i+\frac{p\,^i}{\omega {p}^
2}\,\left(\sqrt{\epsilon m^2+{p}^{2}\omega^2 t_0^2}-\sqrt{\epsilon m^2+{p}^2
\omega^2 t^2}\, \right)\,,
\end{equation}
of a massive particle passing through the point $\vec{x}_0$ at time $t_0$. The light-like case must be treated separately finding that the trajectory of a massless particle reads
\begin{equation}
{x}^i(t)={x}_0^i+n^i\left(t_0-t \right)\,,
\end{equation}
where the unit vector $\vec{n}$ gives the propagation direction.

{ \begin{figure}
		\centering
		\includegraphics[scale=0.24]{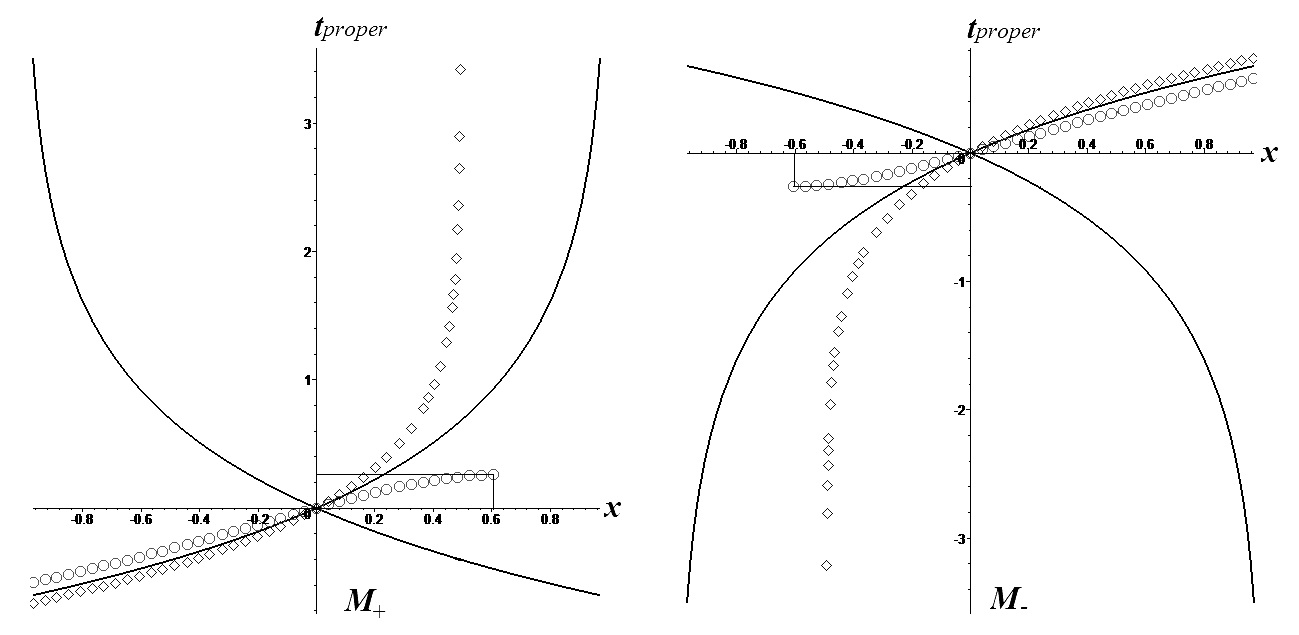}
		\caption{The worldlines of the tardyons  (diamonds) and tachyons (circles) with the same momentum and the initial conditions $x_0=0$ and either $t_0=-\frac{1}{\omega}$ on $M_+$ (left panel) or $t_0=\frac{1}{\omega}$ on $M_-$ (right panel).  The solid lines represent the light-cones which tends asymptoticly to the event horizon at $x=\pm\frac{1}{\omega}$.}
	\end{figure}}

Among the conserved quantities that can be derived as in Ref. \cite{CCC} the energy of the massive particles,
\begin{equation}
E=\omega\, \vec{x}_0\cdot \vec{p}+\sqrt{\epsilon m^2+{p}^{2}\omega^2
t_0^2}\,,
\end{equation}
indicates the allowed domains of our parameters. Thus we see that the tardyonic  particles can have any momentum, $\vec{p}\in {\Bbb R}^3_p$, since their energies remain always  real numbers.  When the  tardyonic particle is at rest, staying in $\vec{x}(t)=\vec{x}_0$ with $\vec{p}=0$, then we find the same rest energy $E_0=m$  as in special relativity.  Thus we conclude that the tardyonic particles behave familiarly just as in Minkowski flat spacetime. 

However, the tachyons have a new and somewhat strange behavior. When $\epsilon=-1$ there are real trajectories only for large momentum with $|\vec{p}|>m$, corresponding to the superluminal motion. Moreover, we must have $p^2\omega^2 t_0^2\ge m^2$ which means that $t_0$ satisfies either the condition $t_0\le -\frac{m}{p\omega}$ on $M_+$ or the symmetric one $t_0\ge \frac{m}{p\omega}$ on $M_-$. Thus we conclude that a classical tachyon of momentum $p>m$  can exist on the expanding portion only in the  time domain $-\infty<t\le-\frac{m}{p\omega}$ while on the collapsing portion its time domain is  $ \frac{m}{p\omega}\le t<\infty$.  The corresponding  time domains of the tachyons with momentum $p>m$  in the charts of proper time are now $(-\infty,\frac{1}{\omega}\ln(\frac{p}{m}))$ on $M_+$ and $(-\frac{1}{\omega}\ln(\frac{p}{m}), \infty)$ on $M_-$. In Fig. (1) we represent the worldlines  in terms of the proper time with the  initial conditions  read   $x_0=0$  at $t_0=\mp\frac{1}{\omega} $ for which $t_{proper}=0$ on both the portions of $M$. 

These restrictions upon the tachyonic lifetime are quite unusual opening the problem of finding a plausible mechanism explaining the tachyon death  on $M_+$ or how this is born  on $M_-$.

\section{Quantum modes}

The next step is to verify if similar restrictions could could arise  in the case of the quantum fields too. For this purpose we have to analyse the tachyonic solutions of the scalar and Dirac fields in momentum representation.

 The specific feature of the quantum mechanics on $M$ is that the energy
operator does not commute with the components of the momentum operator. Therefore,   the  energy and momentum cannot be measured simultaneously with a desired accuracy.
Consequently, there are no particular solutions of the Klein-Gordon or Dirac equations with
well-determined energy and momentum, being forced to consider different plane
waves solutions depending either on momentum or on energy and momentum
direction. In what follows we restrict ourselves to the plane waves of determined momentum.

\subsection{Scalar plane waves}

In an arbitrary chart $\{ x\}$  the action of a charged scalar field $\phi$ of
mass $m$, {\em minimally} coupled with the gravitational field, reads
\begin{equation}\label{action}
{\cal S}[\phi,\phi^*]=\int d^4 x \sqrt{g}\,{\cal L}=\int d^4 x
\sqrt{g}\left(\partial^{\mu}\phi^*\partial_{\mu}\phi-\epsilon m^2 \phi^* \phi\right)\,,
\end{equation}
where $g=|\det(g_{\mu\nu})|$ and $\epsilon$ is our parameter which gives the tardyonic, tachyonic or light-like behaviour. This action gives rise to the KG equation
\begin{equation}\label{KG}
\frac{1}{\sqrt{g}}\,\partial_{\mu}\left[\sqrt{g}\,
g^{\mu\nu}\partial_{\nu}\phi\right]+\epsilon m^2\phi=0\,,
\end{equation}
whose solutions have to be normalized (in generalized sense) with respect to 
the relativistic scalar product \cite{BD},
\begin{equation}\label{SP1}
\langle \phi,\phi'\rangle=i\int_{\Sigma} d\sigma^{\mu}\sqrt{g}\, \phi^*
\stackrel{\leftrightarrow}{\partial_{\mu}} \phi'\,,
\end{equation}
written with the notation $f\stackrel{\leftrightarrow}{\partial}h=f(\partial g)-g(\partial f)$.

In the chart $\{t,\vec{x}\}$ we use here the  Klein-Gordon equation takes the form,
\begin{equation}\label{KG1}
\omega^2 t^2\left( \partial_t^2-\frac{2}{t}\partial_t-\Delta \right)\phi(x)+\epsilon m^2 \psi(x)=0\,.
\end{equation}
The solutions of this equation may be square integrable functions or tempered
distributions with respect to the scalar product (\ref{SP1}) that for
$\Sigma={\Bbb R}^3$ takes the form
\begin{equation}\label{SP2}
\langle \phi,\phi'\rangle=i\int d^3x\, e^{3\omega t}\, \phi^*(x)
\stackrel{\leftrightarrow}{\partial_{t}} \phi'(x)\,.
\end{equation}

It is known that the KG equation (\ref{KG1}) of NP can be analytically solved
in terms of Bessel functions \cite{BD}. There are fundamental solutions of positive frequencies and  given momentum, $\vec{p}$,  that read
\begin{equation}\label{fp}
f_{\vec{
p}}(x)=\frac{1}{2}\sqrt{\frac{\pi}{\omega}}\frac{e^{-\frac{1}{2}i\pi k}}{(2\pi)^{3/2}}\,(-\omega
t)^{\frac{3}{2}}H^{(1)}_{k}\left(-pt\right) e^{i \vec{ p}\cdot \vec{x}}\,,
\end{equation}
where $H^{(1)}_{\nu}$ are Hankel functions, $p=|\vec{p}|$ and we denote
\begin{equation}\label{k}
k=\sqrt{\frac{9}{4}-\epsilon \frac{m^2}{\omega^2}}\,. 
\end{equation}
Obviously, the fundamental solutions of negative frequencies are $f_{\bf p}^*(x)$. 

Now we observe that the only parameter depending on $\epsilon$ is just that given by Eq. (\ref{k}) which encapsulates the information about the nature of the scalar particle. We observe first that there are no major differences between the tardyonic ($\epsilon=1$) and tachyonic ($\epsilon=-1$) cases. The only property depending on the particle's nature is the behavior of the mode functions in the limit of $t\to 0$, when the proper time tends to $\infty$ on $M_+$ or to $-\infty$ on $M_-$ as in Eq. (\ref{ttpr}). The functions (\ref{fp}) are finite in the considered limit only if we have $k\le\frac{3}{2}$ which holds only for  $\epsilon=1$. Thus we find that the tardyonic mode functions remain finite for $t\to 0$ while the tachyonic ones diverge in this limit. 

However, this is not an impediment as long as the conserved scalar product and the conserved quantities do not depend explicitly on the functions giving the time modulation.
Tus we can say that the quantum scalar tachyons may live on $M$ at any time. 

\subsection{Dirac plane waves}

The  tardyonic free Dirac field $\psi$ of mass $m$ and minimally coupled to the gravity of $M$ has the action
\begin{equation}\label{actionD}
{\cal S}[\psi]=\int d^4 x\sqrt{g}\, \left({\cal L}_D(\psi)-m\overline{\psi}\psi\right)\,.
\end{equation}
where the massless Lagrangian density,
\cite{CD1},
\begin{equation}
{\cal L}_D(\psi)=
\frac{i}{2}\,[\overline{\psi}\gamma^{\hat\alpha}D_{\hat\alpha}\psi-
(\overline{D_{\hat\alpha}\psi})\gamma^{\hat\alpha}\psi]\,,\quad \overline{\psi}=\psi^+\gamma^0\,,
\end{equation}
depends on  the covariant derivatives in local frames, $D_{\hat\alpha}$ \cite{CD1}, that guarantee the tetrad-gauge invariance. The point-independent Dirac matrices $\gamma^{\hat\mu}$ satisfy $\{ \gamma^{\hat\alpha},\, \gamma^{\hat\beta} \}=
2\eta^{\hat\alpha \hat\beta}$ giving rise to the basis-generators
$S^{\hat\alpha \hat\beta}=i [\gamma^{\hat\alpha}, \gamma^{\hat\beta}
]/4$ of the spinor representation $(\frac{1}{2},0)\otimes (0,\frac{1}{2})$ of the $SL(2,\Bbb C)$ group that induces the spinor covariant representations \cite{CD1}. 

The Lagrangian theory of the Dirac tachyons \cite{DTach,DTach1,DTach2} allows us to introduce our parameter $\epsilon$ for studying simultaneously the tardyonic, neutrino and tachyonic cases.  This can be done on a natural way considering the new action
\begin{equation}\label{Se}
{\cal S}_{\epsilon}[\psi_L,\psi_R]=\int d^4 x\sqrt{g}\,\left[{\cal L}_D(\psi_L)+\epsilon{\cal L}_D(\psi_R)-\epsilon m(\overline{\psi}_L\psi_R+\overline{\psi}_R\psi_L)\right]\,,
\end{equation}
depending on the chiral projections $\psi_L=L\psi$ and $\psi_R=R\psi$ given by the standard projectors
\begin{equation}
L=\frac{1-\gamma^5}{2}\,, \quad R=\frac{1+\gamma^5}{2}\,.
\end{equation}
Note that this action is written in the style of the Standard Model such that the left-handed term is independent on $\epsilon$ in order to do not affect the  $SU(2)_L$  gauge symmetry.

The action (\ref{Se}) gives the field equations of the massive fields,
\begin{equation}
\begin{array}{lll}
\epsilon=1&(i\gamma^{\hat\alpha}D_{\hat\alpha}-m)\psi=0&{\rm tardyon}\\
&&\\
\epsilon=-1&(i\gamma^5\gamma^{\hat\alpha}D_{\hat\alpha}+m)\tilde \psi=0&{\rm tachyon}
\end{array}
\end{equation}
while for $\epsilon=0$ we recover the usual theory of the massless neutrino. It is remarkable that these equations are related through
\begin{equation}\label{tata}
\tilde \psi(x,m)=\tau \psi(x,im)
\end{equation}
where the unitary matrix
\begin{equation}
\tau=\frac{1}{\sqrt{2}}(1-i\gamma^5)
\end{equation}
has the properties $\tau^+=\tau^{-1}$, $\overline{\tau}=\tau$ and $\tau^2=-i\gamma^5$. This means that it is enough to solve the tardyonic equation for finding simultaneously the tachyonic solution by using Eq. (\ref{tata})

Let us start with the tardyonic case in the chart $\{t,\vec{x}\}$ and tetrad-gauge 
\begin{equation}\label{tt}
e^{0}_{0}=-\omega t\,, \quad e^{i}_{j}=-\delta^{i}_{j}\,\omega t
\,,\quad
\hat e^{0}_{0}=-\frac{1}{\omega t}\,, \quad \hat e^{i}_{j}=-\delta^{i}_{j}\,
\frac{1}{\omega t}\,.
\end{equation}
where the free Dirac equation for tardyons reads \cite{CD1},
\begin{equation}\label{EDir}
\left[-i\omega t\left(\gamma^0\partial_{t}+\gamma^i\partial_i\right)
+\frac{3i\omega}{2}\gamma^{0}-m\right]\psi(x)=0\,.
\end{equation}
The general solution allows the mode expansion in momentum representation,
\begin{equation}\label{psiab}
\psi(t,\vec{x})=\int d^3 p \sum_{\sigma}\left[U_{\vec{p},\sigma}(x)
a(\vec{p},\sigma)+V_{\vec{p}, \sigma}(x){a^c}^{\dagger}(\vec{p},
\sigma) \right]\,,
\end{equation}
written in terms of the field operators, $a$ and $a^c$, and  the particle and antiparticle fundamental spinors of this basis, $U_{\vec{p},\sigma}$ and respectively $V_{\vec{p},\sigma}$,  which depend on the momentum $\vec{p}$ (with $p=|\vec{p}|$) and polarization $\sigma=\pm 1/2$. According to our prescription of frequencies separation on the expanding portion (of the Bounch-Davies type) we find that these spinors,  in the standard representation of the Dirac matrices (with diagonal $\gamma^0$), take the form \cite{CD1},
\begin{eqnarray}
U_{\vec{p},\sigma}(t,\vec{x}\,)&=& i N (\omega t)^2\left(
\begin{array}{c}
e^{\frac{1}{2}\pi\mu}H^{(1)}_{\nu_{-}}(-p t) \,
\xi_{\sigma}\\
e^{-\frac{1}{2}\pi\mu}H^{(1)}_{\nu_{+}}(-p t) \,\frac{\vec{\sigma}\cdot\vec{p}}{p}\,\xi_{\sigma}
\end{array}\right)
e^{i\vec{p}\cdot\vec{x}}\label{Ups}\\
V_{\vec{p},\sigma}(t,\vec{x}\,)&=&-i N (\omega t)^2 \left(
\begin{array}{c}
e^{-\frac{1}{2}\pi\mu}H^{(2)}_{\nu_{-}}(-p t)\,\frac{\vec{\sigma}\cdot\vec{p}}{p}\,
\eta_{\sigma}\\
e^{\frac{1}{2}\pi\mu}H^{(2)}_{\nu_{+}}(-p t) \,\eta_{\sigma}
\end{array}\right)
e^{-i\vec{p}\cdot\vec{x}}\,.\label{Vps}
\end{eqnarray}
The notation $\sigma_i$ stands for the Pauli matrices while $H_{\nu_{\pm}}^{(1,2)}$  are the Hankel functions of indices $\nu_{\pm}=\frac{1}{2}\pm i\mu$ with $\mu=\frac{m}{\omega}$. The normalization constant $N$ has to be calculated according to a normalization condition on $M_+$  that will not be discussed here. Similar solutions can be obtained on $M_-$ by changing $\omega \to -\omega$.

Now we can write directly the tachyonic fundamental solutions using Eq. (\ref{tata}). We obtain the final result as
\begin{eqnarray}
&&\tilde U_{\vec{p},\sigma}(t,\vec{x}\,)\nonumber=\\
&& i \tilde N (\omega t)^2\left[
\begin{array}{c}
\left(e^{\frac{i}{2}\pi\mu}H^{(1)}_{\tilde\nu_{+}}(-p t)-ie^{-\frac{i}{2}\pi\mu}H^{(1)}_{\tilde\nu_{-}}(-p t) \,\frac{\vec{\sigma}\cdot\vec{p}}{p} \,
\right)\xi_{\sigma}\\
\left(e^{-\frac{i}{2}\pi\mu}H^{(1)}_{\tilde\nu_{-}}(-p t) \,\frac{\vec{\sigma}\cdot\vec{p}}{p}-ie^{\frac{i}{2}\pi\mu}H^{(1)}_{\tilde\nu_{+}}(-p t)\right)\,\xi_{\sigma}
\end{array}\right]
e^{i\vec{p}\cdot\vec{x}}\,,\\
&&\tilde V_{\vec{p},\sigma}(t,\vec{x}\,)=\nonumber\\
&&-i  \tilde N (\omega t)^2 
\left[\begin{array}{c}
\left(e^{-\frac{i}{2}\pi\mu}H^{(2)}_{\tilde\nu_{+}}(-p t)\,\frac{\vec{\sigma}\cdot\vec{p}}{p}\,
-i e^{\frac{i}{2}\pi\mu}H^{(2)}_{\tilde\nu_{-}}(-p t)\right)\eta_{\sigma}\\
\left(e^{\frac{i}{2}\pi\mu}H^{(2)}_{\tilde\nu_{-}}(-p t)-i e^{-\frac{i}{2}\pi\mu}H^{(2)}_{\tilde\nu_{+}}(-p t)\,\frac{\vec{\sigma}\cdot\vec{p}}{p}\right) \,\eta_{\sigma}
\end{array}\right]
e^{-i\vec{p}\cdot\vec{x}}\,,
\end{eqnarray}
where now all the indices $\tilde\nu_{\pm}=\frac{1}{2}\pm \frac{m}{\omega}$ are real numbers. These fundamental spinors are definite on the whole expanding portion without restrictions despite of their complicated form. On the collapsing portion we meet similar solutions but with $\omega \to -\omega$.

We note that, as in the scalar case, there are mode functions that diverge in the limit of $t\to 0$ but only for the tachyonic mass $m>\frac{3}{2}\omega$ while for $m< \frac{3}{2}\omega$ these functions remain finite as in the tardyonic case. This situation is different from the scalar case where all the tachyonic scalar functions are divergent in this limit. However, this phenomenon has no physical consequences since only the conserved quantities (scalar products and expectation values) have physical significance.  

Thus we can say that the Dirac tachyons on the de Sitter background have fundamental solutions well-defined on the entire physical space at any moment.

\section{Conclusion}

Finally, we must stress that the tachyon physics on the de Sitter manifold seems to lead to a  fundamental contradiction between the classical and quantum cases. Thus in the classical approach  the trajectory  must end when the energy becomes imaginary. On the contrary, in the quantum theory there are no time restrictions upon the mode functions of the free fields that behave as tempered distributions on the whole physical space at any moment. 

However, in our opinion it is premature to draw definitive conclusions concerning this discrepancy before studying the propagation and dispersion of the wave packets and the conserved quantities which could offer new arguments in what concerns the relation between the classical and quantum tachyons. We expect that one may face to serious difficulties in this domain but that could be overdrawn by using new analytical or even numerical methods \cite{Mitica}.

\section*{Acknowledgments}

This paper  is supported by a grant of the Romanian National Authority for Scientific Research,
Programme for research-Space Technology and Advanced Research-STAR, project nr. 72/29.11.2013 between Romanian Space Agency and West University of Timisoara.

\end{document}